\documentclass[10pt,aps,prb,twocolumn,superscriptaddress]{revtex4}

\usepackage{graphicx}
\usepackage{latexsym}

\begin{document}
\bibliographystyle{unsrt}

\title{Cicada's wings as determinant factor for the sound emission: The case of \textit{Quesada gigas}.}

\author{C. L. De Santis}
\affiliation{Facultad de Ingenier\'{\i}a Qu\'{\i}mica and Facultad
de Humanidades y Ciencias, Universidad Nacional del Litoral,
Santiago del Estero 2829 3000 Santa Fe Argentina.}

\author{R. Urteaga}
\affiliation{Laboratorio de Cavitaci\'{o}n y Biotecnolog\'{\i}a.
Centro At\'{o}mico Bariloche, Avda Bustillo km 9.5, 8400 San
Carlos de Bariloche, R\'{\i}o Negro, Argentina.}

\author{P. G. Bolcatto}
\affiliation{Facultad de Ingenier\'{\i}a Qu\'{\i}mica and Facultad
de Humanidades y Ciencias, Universidad Nacional del Litoral,
Santiago del Estero 2829 3000 Santa Fe Argentina.}

\date{\today}


\begin{abstract}
Cicadas (Homoptera:Cicadidae) are insects able to produce loudly
songs and it is known that the mechanism to produce sound of
tymballing cicadas works as a Helmholtz resonator. In this work we
offer evidence on the participation of the wings in a high quality
resonating process which defines the details of the acoustic
properties of the calling song. The study is carry on
\textit{Quesada gigas} species and it is dividied in three stages:
(i) the acoustical characterization of the abdominal cavity, (ii)
the record and calculation of frequency spectrum of the calling
song, and (iii) the measurement of the vibration modes of the
wings. The comparison between all the results unequivocally show
the dramatic influence of the wings in the moment in which the
insect emits its calling song.
\end{abstract}


\maketitle

\section{Introduction}
Cicadas is the generic name of more than 1500 species of insects
belonging to the Homoptera order, Cicadidae family
\cite{klots:1960}. Distinguishing them from other insects, cicadas
are able to accomplish collective behaviors
\cite{greenfield:1994,sueur:2002,sueuraubin:2002} and rich
communicational codes \cite{cooley:2001,stolting:2002}. In
particular, males cicadas have modified organs to produce sounds
in order to attract conspecific females and to prevent predators
\cite{barnes:1996,walker:2004}. It is well-accepted that the
sound-generator apparatus of typical cicadas works as a Helmholtz
(bottle-shaped) resonator \cite{young:1990}. However, the calling
song of cicadas is a species-specific sound that shows complex
frequency spectra which evidence the participation of other
mechanisms in the sound production. This is the case of
\textit{Proarna dactyliophora} Berg, \textit{Proarna bufo}
Distant, \textit{Dorisiana drewseni} (St\aa l), \textit{Dorisiana
viridis} (Olivier) and \textit{Quesada gigas} Olivier, species
occurring in Santa Fe city (Argentina) and its outskirts ($31^{o}
38'$ S latitude, $60^{o} 42'$ W longitude)
\cite{desantis:2003,desantis:2005}. In this work we are going to
focuss on the case of \textit{Q. gigas}, the only species with a
broad distribution covering North to South America
\cite{moore:1993}. They have a high size reaching a corporal
length of $(43 \pm 4)$mm (mean $\pm$ s.d.) \cite{desantis:2003}
and its song is distinctive from other species since it sounds as
a "whistle". Likewise that in the rest of typical or tymballing
cicadas, its sound-generator apparatus is composed by a pair of
stiffened membranes, the tymbals, placed dorsolaterally in the
first abdominal segment. Two muscles are attached to the tymbals,
the so-called tymbal and the tensor muscles. The tymbal muscle has
great dimensions and due to strong contractions buckles the
membrane inwards for then to recover the initial situation because
of the elastic energy stored. The tensor muscle does not work in
opposition to tymbal muscle but it changes the elastic properties
of the tymbal membrane and can modify the amplitude, the time
interval or frequency of the acoustic signal \cite{hennig:1994}.
The generated pulse is amplified by the almost hollow abdominal
cavity, and radiated by the tympana \cite{young:1990}. Understood
in this way, the sound-producing apparatus of tymballing cicadas
is a pulsed Helmholtz resonator in which the abdominal sac is the
resonant cavity, the tympanal opening is the neck and two
cuticular flaps, the opercula, regulate the neck area
\cite{bennet-clark:1992,bennet-clark:1999}. The predominant
frequency of a Helmholtz resonator is given by the geometry of the
system \cite{morse} or as in this case, by the morphometric
features of each species. Thus, it is possible to find a scaling
law correlating the frequency of the song and some typical length
\cite{bennet-clark:1994}. However, the scaling law fails in some
species as is the case of \textit{Q. gigas} where the predicted
calling frequency is roughly 4 kHz while the actual predominant
frequency is the order of 1 kHz \cite{desantis:2003}.

In this paper we will show how the wings of \textit{Q. gigas}
species dramatically participate in a high-quality resonant
process defining its calling song being this, up to our knowledge,
the first evidence about cicada's wings as determinant factor in
sound production.

\section{Methods}
In order to reach the objectives of this work, we proceed in three
stages: (i) Firstly we characterized the response in frequencies
of the abdominal cavity, then (ii) we recorded the calling song of
\textit{Q. gigas} and we calculated its frequency spectra, (iii)
we measured the normal vibration modes of the wings and finally,
we compared all the results to arrive to the final conclusion.

The physical characterization of the abdominal cavity as
acoustical resonator should be done extracting the response in
frequency after exciting it with a brief pulse. Experimentally we
are able to obtain this response taking advantage of the
information given by the distress or protest call. In our case,
this song was recorded holding the animal with the hand and moving
the wings towards a side.

The calling songs were recorded in the natural habitat of this
species at different light hours, temperatures greater than 30 C
and covering urbanized and non-urbanized areas of Santa Fe city,
Argentina.

Both, distress and calling song were recorded with SONY TM-343 and
Panasonic RQ-L309 magnetic tape recorders, provide with electret
condenser microphones which guarantee a planar response in the
frequency range of interest. Then, the records were digitalized
with a rate of 22050 data per second.

\begin{figure}[htb]
\includegraphics[width=0.4\textwidth]{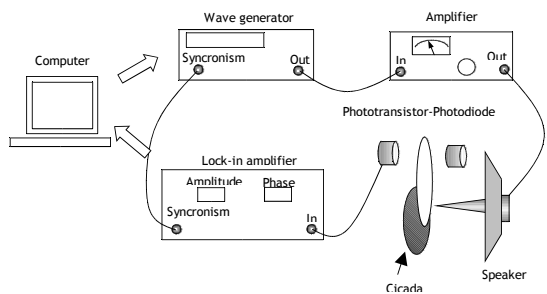}
\caption{\textbf{Experimental setup to measure the vibrational
normal modes of \textit{Q. gigas} wings}. The light cone attached
to the speaker stimulate by outside the wings with a controlled
frequency just in the region where the tymbal membrane is placed.
The wing's movement produces a shadow pattern over the light
detector and the Lock-in amplifier guarantees the measurement of
the amplitude of oscillation for the same frequency of the signal
generator.} \label{Fig1}
\end{figure}

The experimental setup used to measure the vibration modes was
designed to detect the wing movement. It is schematically shown in
Fig. \ref{Fig1}.  A speaker governed by a signal generator (15MHz
Arbitrary waveform generator HP 33120A) produces a controlled
stimulus and a light cone attached to it localizes the stimulus
over the wing just in the region where the tymbal membrane is
placed. When the external frequency coincides with a normal
vibrational mode, the wing's movement is increased and the
intensity of the light beam over the photoreceptor changes. A
Lock-in amplifier (Stanford Research Systems SR530) receives the
signal and the measurement is done in synchrony with the external
frequency. The measurements were taken upon the same individual
from which the distress call and calling song were recorded, but
in this case the animal was dead.

\begin{figure}[htb]
\includegraphics[width=0.3\textwidth]{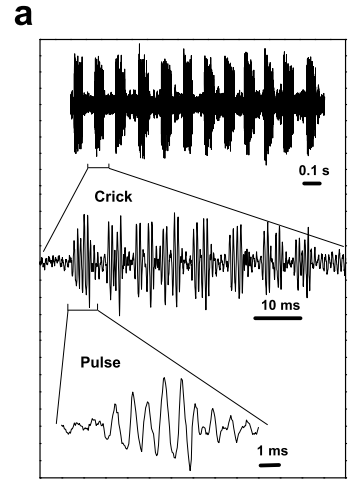}
\includegraphics[width=0.3\textwidth]{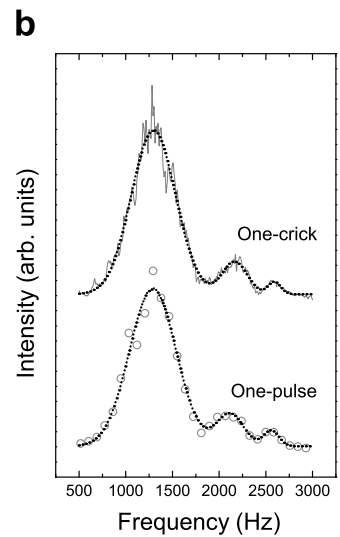}
\caption{\textbf{Physical characterization of the abdominal cavity
of \textit{Q. gigas}}. (a) Oscillogram of distress call and
details of crick and pulse. Records were digitalized with a rate
of 22050 data per second. (b) Response function of the abdominal
cavity resonator. The results correspond to an average on Fourier
spectra $(n=50)$ of cricks and first pulses of the distress call.}
\label{Fig2}
\end{figure}

\section{Results and Discussion}
The adopted procedure to record the distress song assure us that
the emitted sound is an unique feature of the pulsated of the
muscle and tymbal membrane over the abdominal cavity. An example
of this record is shown in Fig. \ref{Fig2}a. The upper oscillogram
show a series of brief emissions that we will name by the
onomatopoeic word "crick". Each crick in turn, is composed by a
train of 6-10 pulses. We ascribe each pulse to a consecutive
contraction of the muscles buckling the tymbal membrane and its
ribs so that this system works as a sort of Zeeman machine
\cite{zeeman} in virtue of which it is possible to excite
frequencies of the order of the kHz. Therefore, by performing the
Fourier Analysis of the first pulse of each crick we can obtain
the acoustical response function of the abdominal cavity. Figure
\ref{Fig2}b shows the frequency spectrum characterizing the cavity
after having averaged 50 Fourier spectra corresponding to
different cricks emitted by the same individual. Pulse and crick
spectra are quite similar and a three-gaussian fit yields a main
peak at $(1290 \pm 10)$ Hz with a half-height width of $(500 \pm
30)$ Hz for the pulse and $(1296 \pm 3)$ Hz and $(478 \pm 6)$ Hz
for the crick respectively.

\begin{figure}[htb]
\includegraphics[width=0.5\textwidth]{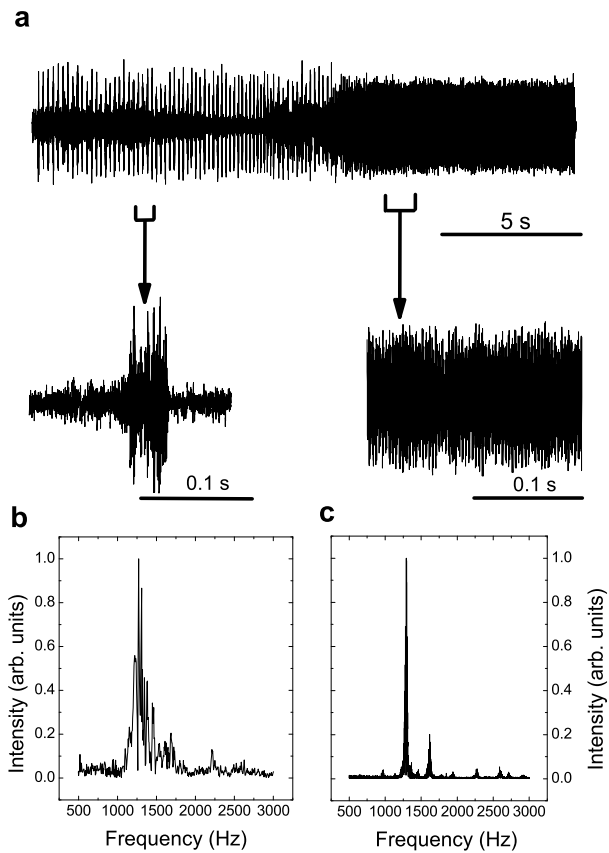}
\caption{\textbf{Calling song of \textit{Q. gigas}}. (a)
Oscillogram of the complete song and details of a crick previous
to the "whistle" sound, and of the sustained phase of the song.
(b) and (c) Fourier spectrum of the crick and sustained sound
respectively.} \label{Fig3}
\end{figure}

On the other hand, the response in time of each pulse lasts $(6.5
\pm 0.3)$ ms $(n=442)$, which allows us to infer that if the
calling song of cicadas was a continuous sequence of pulses
resonating into the abdominal cavity, the corresponding Fourier
spectrum should be simply a set of peaks separated by $(154 \pm
7)$ Hz approximately and modulated by the one-pulse envelope
function given by Fig. \ref{Fig2}b. Nevertheless, the sustained
phase of the calling song of \textit{Q. gigas} does not agree with
this conclusion as it can be observed in Fig. \ref{Fig3}. The
complete calling song lasts around 15-20 s and it is composed by
three phases. Firstly, the animal emits a sequence of cricks, then
occurs a short and unstable phase which precedes the sustained
song. Figure \ref{Fig3}a is the complete oscillogram with details
of the first and third phases and Figs. \ref{Fig3}b and
\ref{Fig3}c are their corresponding Fourier spectra. The sustained
part of the song involves a very well defined spectrum with a main
peak around $(1294 \pm 1)$ Hz, which is in good agreement with the
main peak of the abdominal cavity. However, the width of this very
thin peak is more than ten times lower than the corresponding to
the Helmholtz-like resonator [$(29 \pm 1)$ Hz in this particular
record] (Fig. \ref{Fig3}c). Besides, there are noticeable
differences in lineshape between cricks corresponding to the
calling song (Fig. \ref{Fig3}b) and protest song (Fig.
\ref{Fig2}b). Therefore, an evident conclusion arising from the
comparison among Figs. \ref{Fig2}b, \ref{Fig3}b and \ref{Fig3}c is
that, with the purpose to produce the sustained song, \textit{Q.
gigas} adds a new mechanism in its sound-generator apparatus.

\begin{figure}[htb]
\includegraphics[width=0.4\textwidth]{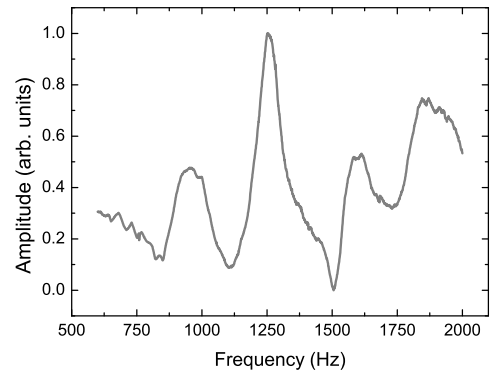}
\caption{\textbf{Measured amplitude spectrum for the wing
movement} using the experimental setup of Fig. \ref{Fig1}. The
main peak is located at 1257 Hz nearly coincident with the main
peak of distress and calling song.} \label{Fig4}
\end{figure}

Inspecting the morphology of tymballing cicadas, one may verify
that the tymbal membrane is located dorsolaterally but clearly
exposed to the outside and covered and in contact with the wings.
This biological design notice us about the unavoidable interaction
between the buckling membrane and the wings when the calling song
is emitted. In order to quantify the relevance of this interaction
we measured the normal modes of vibration of \textit{Q. gigas}
wings and we compared these results with the corresponding
recorded songs. The amplitude spectrum is drawn in Fig.
\ref{Fig4}. In the region of frequencies of interest (600-2000 Hz)
four peaks are discernible. The central peak is located at 1257
Hz, nearly coincident with the maximum of calling and distress
song, and the lateral peaks are placed at 959 Hz, 1601 Hz, and
1878 Hz respectively.

\begin{figure}[htb]
\includegraphics[width=0.4\textwidth]{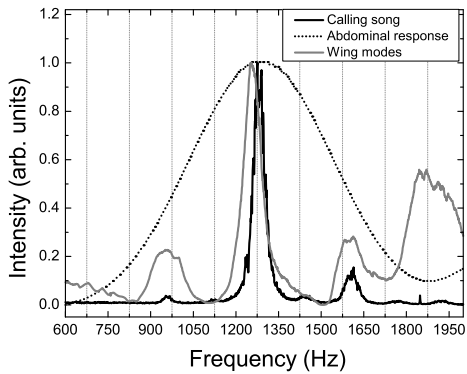}
\caption{\textbf{Overall comparison} between calling song (black
line), acoustical response of abdominal cavity (dotted curve) and
experimental normal modes (intensity spectrum) of the wings of
\textit{Q. gigas} (grey line). All the spectra are normalized by
fixing their maximum value equal to one.} \label{Fig5}
\end{figure}

In Fig. \ref{Fig5} we summarize a comparison among spectra
corresponding to the pulse of distress call, the calling song and
the square of the amplitude spectra of the wing's vibration modes
since it is the quantity proportional to the acoustical power
signal indeed. It had been mentioned that if the calling song was
only the consequence of a regular pulsed over the abdomen, its
Fourier spectrum should be represented by equidistant peaks
modulated by the one-pulse envelope curve. By contrast, if we
assume an interaction between wings, tymbal and abdomen, we can
infer that deeps at 1110 Hz and 1508 Hz in the wing spectrum
depress the corresponding peaks that should appear due to the
beating. Moreover, the slight asymmetry of the main peak in the
intensity wing spectrum allows the occurring of a small peak at
1440 Hz in the calling song. In this way, the interplay among
tymbal, abdomen and wings produces an energy optimization over the
main peak defining the sustained part of the calling song of
\textit{Q. gigas}.

\begin{figure}[htb]
\begin{center}
\includegraphics[width=0.5\textwidth]{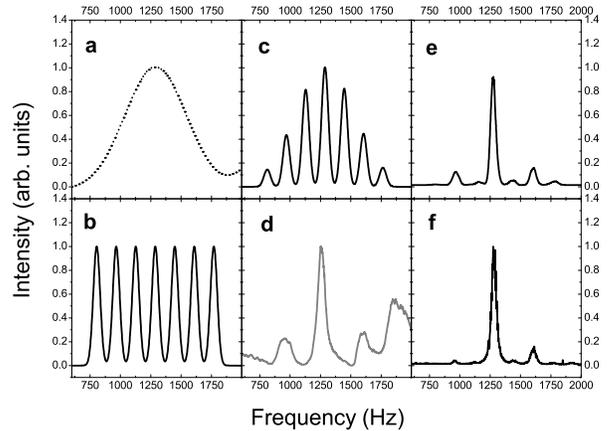}
\end{center}
\caption{\textbf{Reconstruction of the calling song spectrum of
\textit{Q. gigas}}. (a) Abdominal acoustic response. (b)
Simulation of the tymbal beating using gaussian functions with a
half-width of 40 Hz and regularly separated by 160.9 Hz. (c)
modulated beating arising from the product of (a) and (b) curves.
(d) Intensity spectrum of the wing normal modes. (e) Simulated
calling song frequency spectrum reached by multiplying (c) and (d)
curves. (f) Real calling song spectrum.} \label{Fig6}
\end{figure}

Fig. \ref{Fig6} includes additional analysis enforcing the
comprehension of the wing-abdomen-tymbal coupled resonator forming
the sound-producing system. Fig. \ref{Fig6}c is the product
between the frequency response of the pulse (Fig. \ref{Fig6}a) and
an analytical function simulating an assumed regular pulsed of the
tymbal (Fig. \ref{Fig6}b). Now, by multiplying this curve by the
intensity spectrum of the wing's vibrational modes (Fig.
\ref{Fig6}d) it is obtained the curve shown in Fig. \ref{Fig6}e,
which is very much similar to the calling song spectrum (Fig.
\ref{Fig6}f). In conclusion, the abdomen morphometry (represented
by Fig. \ref{Fig6}a), the tymbal beating (simulated by the
function in Fig. \ref{Fig6}b) and the geometry and stiffness of
the wings (signed by the Fig. \ref{Fig6}d) strongly interact in
order to define the acoustic feature of the calling song of
species \textit{Q. gigas}.

\section{Conclusions}
The Fourier spectra of the sound emitted by cicadas can be used as
taxonomic tool due to the clear differences between species. In
the species \textit{P. dactyliophora}, \textit{P. bufo},
\textit{D. drewseni}, \textit{D. viridis} and \textit{Q. gigas},
occurring in Santa Fe, Argentina, we had previously verified that
the overall explanation of their frequency spectra can not be
performed only with the identification of the insect as a
Helmholtz-like acoustic machine. In this work we have found novel
mechanisms in the sound production in cicadas. While not strictly
a Helmholtz resonator, the abdominal cavity is actually the
resonator which defines the value of the predominant frequency of
the calling song. The quality of the sound, however is a
consequence of a very fine interaction between the beating tymbal
membrane and the wings, involving a selection of frequencies which
allows to the species \textit{Q. gigas} to produce an almost
musical sound in its calling song.

\begin{acknowledgements} R. U. and P. G. B. acknowledge the
financial support by Consejo Nacional de Investigaciones
Cient\'{\i}ficas y Tecnol\'{o}gicas (CONICET). This work was
supported by Grant 989/04 CONICET. The authors acknowledge to
Javier Schmidt for the critical reading of the manuscript.
\end{acknowledgements}

\bibliography{bibbio}

\end{document}